\documentclass[%
 reprint,
superscriptaddress,
 amsmath,amssymb,
 aps,
prl,
]{revtex4-2}

\usepackage{graphicx}
\usepackage{dcolumn}
\usepackage{bm}
\usepackage{cleveref}
\usepackage{physics}
\usepackage{dsfont}
\usepackage[normalem]{ulem}
\usepackage{xcolor}
\usepackage[makeroom]{cancel}

\newcommand{\jls}[1]{\textcolor{magenta}{\ifmmode\text{\cancel{\ensuremath{#1}}}\else\sout{#1}\fi}}

\begin{document}

\preprint{APS/123-QED}

\title{The role of atomic interactions in cavity-induced continuous time crystals}

\author{Christian H. Johansen}
 \affiliation{Max-Planck-Institut f\"ur Physik komplexer Systeme, 01187 Dresden, Germany}
\author{Johannes Lang}%
\affiliation{Institut f\"ur Theoretische Physik, Universit\"at zu K\"oln, Z\"ulpicher Stra{\ss}e 77, 50937 Cologne, Germany}%
\affiliation{Max-Planck-Institut f\"ur Physik komplexer Systeme, 01187 Dresden, Germany}%
\author{Francesco Piazza}
\affiliation{Theoretical Physics III, Center for Electronic Correlations and Magnetism,
Institute of Physics, University of Augsburg, 86135 Augsburg, Germany}
\affiliation{Max-Planck-Institut f\"ur Physik komplexer Systeme, 01187 Dresden, Germany}%

\date{\today}

\begin{abstract}
We consider continuous time-crystalline phases in dissipative many-body systems of atoms in cavities, focusing on the role of short-range interatomic interactions. 
First, we show that the latter can alter the nature of the time crystal by changing the type of the underlying critical bifurcation. Second, we characterize the heating mechanism and dynamics resulting from the short-range interactions and demonstrate that they make the time crystal inherently metastable. We argue that this is generic for the broader class of dissipative time crystals in atom-cavity systems whenever the cavity loss rate is comparable to the atomic recoil energy. We observe that such a scenario for heating resembles the one proposed for preheating of the early universe, where the oscillating coherent inflation field decays into a cascade of exponentially growing fluctuations.
By extending approaches for dissipative dynamical systems to our many-body problem, we obtain analytical predictions for the parameters describing the phase transition and the heating rate inside the time-crystalline phase. 
We underpin and extend the analytical predictions of the heating rates with numerical simulations.     
\end{abstract}

\maketitle

\emph{Introduction.---}
Following the first conceptualization of time-crystalline phases of matter \cite{WilczekTC1,WilczekTC2}, it was quickly proven that such phases cannot appear in thermal equilibrium \cite{Nozières2013noTC,Bruno2013noTC,Watanabe2015NoTC}. 
However, it turned out to be possible to realize such phases in periodically driven systems, both closed \cite{Zhang2017dtc, Choi2017dtc,Rovny2018dtc,pal2018dtc,randall2021many,mi2022dtc} and dissipative \cite{Kessler2021tc,taheri2022all}.  

Among the latter, systems of atoms in optical cavities have emerged as an ideal platform to realize continuous time-crystalline phases \cite{Kongkhambut2022tc,Dreon2022tc,LiCurrent}, where an effectively time-independent drive of the atomic system is counterbalanced by the loss of photons out of the cavity mirrors. 
In these phases, continuous time-translation invariance is spontaneously broken, and oscillations persist even though the system possesses a macroscopic number of degrees of freedom, among which energy can be redistributed via interactions.

Since the phase space of scattering by cavity-mediated interactions between atoms is limited, due to their long range, redistribution of energy through these processes is inefficient \cite{piazza2014quantum,schutz2014prethermalization,mivehvar2021Review}. However, the intrinsic atomic short-range interactions allow for efficient redistribution of energy among the atoms. 
Indeed, experiments show strong indications that these interactions are one of the main fundamental limiting factors to the measured lifetime of the time crystal \cite{Kessler2021tc}.

Despite their crucial role short-range atomic interactions have not been theoretically investigated so far in a systematic way for continuous time crystals in atom-cavity setups. 
In this work, we undertake this task. Not only do we provide a full picture of the possible destabilization processes 
but we also show that short-range interactions can alter the nature of the time crystal itself.

We consider a simple and experimentally realizable mechanism for the appearance of time-crystalline phases for an interacting BEC coupled to two cavity modes \cite{Johansen2022floquet}. 
By extending approaches for classical non-linear dissipative systems to our many-body problem, we obtain an analytical description of the time crystal in terms of cavity-induced critical bifurcations and show how inter-atomic interactions can modify the nature of the latter. 
Within this approach, we also compute the dependence of the energy-redistribution rates on external parameters and identify the scattering processes responsible for making the time crystal metastable.

The analytical understanding of the results, which we also underpin with numerical analysis, allows for a deep insight into the generic features of the phenomenology beyond the specific model considered and provides orientation for future investigations both in theory and experiment.

\begin{figure}
    \centering
    \includegraphics[width=\columnwidth]{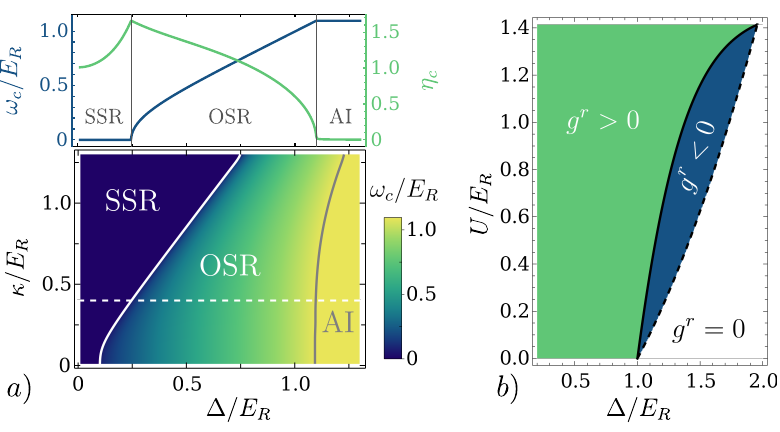}
    \caption{The critical frequency of the instability is shown in the lower plot of a) as a function of $\Delta$ and $\kappa$. By tuning $\Delta$ the critical mode change from exhibiting static to oscillating superradiance and a purely atomic instability over a large range of cavity loss rates. Above the critical frequency and coupling is shown along the white dashed line. The upper plot shows the critical frequency and coupling along the dashed line in the lower plot. 
    In b) the sign of the cubic interaction as a function of $\Delta$ and $U$ is plotted for $\kappa=0.4E_R$. This determines the stability of the symmetry-broken state beyond the linear analysis. For for the entire figure $\delta=0.2 E_R$.
    }
    \label{fig:lcPD}
\end{figure}

\emph{Model.---}
The system considered is an ultracold gas of bosonic atoms in a BEC state, dispersively coupled with equal strength to two modes of an optical cavity. 
In this regime, a photon imparts a recoil momentum of $Q=2\pi/\lambda$ to an atom, with $\lambda$ being the wavelength of the photon in a given mode.
In the thermodynamic limit, the atomic BEC at momentum $k$ is described by a complex field $\psi_k$ satisfying the Gross-Pitaevski mean-field equations. 
Furthermore, in the limit of a small transverse extend of the BEC compared to the cavity waist we can simplify the model to one spatial dimension \cite{mivehvar2021Review,Johansen2022floquet}
\begin{equation}\label{eq:atomEOM}
\begin{aligned}
        i\partial_t\psi_k=&k^2\psi_k+U\sum_{q,q'}\psi_q \psi_{q'} \bar{\psi}_{q+q'-k}\\ &+\frac{\tilde{\eta}}{\sqrt{2}}\sum_{j=1,2} \Re\left(\phi_j\right)\left(\psi_{k+Q}+\psi_{k-Q}\right)\,,
\end{aligned}
\end{equation}
where the bar denotes complex conjugation. 
This equation has been written in units of the recoil energy $E_R=\hbar^2 Q^2/2m$ and in the rotating frame of the laser. 
The time-dependence of the fields is kept implicit and the atom field has been normalized to 1.
The cavity-mode wavelengths have been chosen to be equal, as we assume the modes differ in transverse direction \cite{Johansen2022floquet}. 
The coupling strength $\tilde{\eta}$ can experimentally be tuned by the strength of the transverse pump while the atoms are interacting with each other through a contact interaction of strength $U$. 
The complex field $\phi_j$ corresponds to the coherent cavity-field amplitude which satisfies the equation
\begin{equation}\label{eq:cavityEOM}
    \begin{aligned}
    i\partial_t \phi_j=&\left(\Delta_j-i\kappa\right)\phi_j\\
    &+\frac{\tilde{\eta}}{2\sqrt{2}}\sum_{k=-\infty}^\infty\bar{\psi}_{k}\left(\psi_{k+Q}+\psi_{k-Q}\right),
    \end{aligned}
\end{equation}
where the cavity field has been normalized by the square of the atom number. 
The cavity linewidths, $\kappa$, have been assumed to be identical for both modes. 
In the following we will consider $\kappa$ on an energy scale similar to the recoil energy, as realized for instance in \cite{Kessler2014subrecoil}. 
In the actual implementation of the dispersive atom-cavity coupling, the characteristic frequency of each cavity mode $\Delta_j$ corresponds to the detuning of the mode frequency with respect to laser-driven two-photon transitions \cite{Johansen2022floquet}. 
The steady-state of this model can break time-translation invariance when the two detunings have opposite signs. With this in mind the detunings are parametrized as $\Delta_1=-\left(\Delta-\frac{\delta}{2}\right)$ and $\Delta_2=\Delta+\frac{\delta}{2}$. By choosing $0<\delta<2\Delta$ the negative detuning has the smallest amplitude $\vert\Delta_1\vert<\vert \Delta_2\vert$.   

\emph{Nature of the time crystal.---}
Below a critical coupling strength $\eta_c$, all atoms are in the homogeneous state $\psi_0$, and the coherent part of the cavity fields is empty. 
This configuration is denoted as the normal phase (NP) and it is always a fixed point of the equations of motion \cref{eq:atomEOM,eq:cavityEOM}.   
As $\tilde{\eta}$ is increased beyond $\eta_c$ the NP fixed point becomes unstable and the system enters a state where a fraction of the atom population is transferred to $\psi_{\pm Q}$ and the coherent fields of the cavity becomes finite. 
This symmetry-broken state is often referred to as the superradiant (SR) or self-organized state \cite{Ritsch2002superradiant, Baumann2010Dicke}.
The frequency $\omega_{c}$ of the excitation becoming undamped above $\eta_c$, can be derived through a linear expansion around the NP fixed point \cite{JohansenThesisChp4} (see \cite{del2023limit} for an alternative approach). 
One finds three non-negative real solutions for the frequency of the critical mode. These three solutions are $\omega_c=0$, a resonance at the energy of the Bogoliubov excitation of the BEC at the recoil momentum $\omega_c=\omega_a=\sqrt{E_R\left(E_R+2U\right)}$ and a solution given by
\begin{equation}\label{eq:frequency}
\omega_c= \sqrt{\frac{\delta^2}{4}+\sqrt{\left(4\Delta^2-\delta^2\right)\left(\Delta^2+\kappa^2\right)}-\Delta^2-\kappa^2},
\end{equation}
which is solely determined by cavity parameters, that is, it does not depend on $U$ and $E_R$. 
This feature, which can be attributed to the fact that the cavity is the only dissipation channel, implies a robustness of this self-sustained periodic signal to perturbations of the nonlinear medium that causes this signal to appear in the first place.
Out of the three modes the critical one is identified by having the smallest real critical coupling.
Differently from the frequency, the critical coupling always depends on both cavity and atom parameters (see supplementary) such that the phase diagram will depend on all parameters of the theory. 

In \cref{fig:lcPD}(a) the frequency of the critical mode at $\eta_c$ is plotted as a function of $\kappa$ and $\Delta$ and is a good order parameter for distinguishing the three different phases of the system. 
For $\Delta<\delta/2$ both cavity modes have a positive detuning and the system always exhibits static superradiance (SSR), characterized by a critical mode with zero frequency. 
SSR requires a finite critical atom-cavity coupling such that the critical mode is a polariton.
For $\Delta>\delta/2$ one of the modes acquires a negative detuning. Differently from a positively-detuned mode, a negatively-detuned one disfavors a superradiant density modulation. 
The competition between the two cavity modes induces an oscillating superradiant phase (OSR) \cite{Johansen2022floquet,Mivehvar2022tc}, which also requires a finite coupling strength such that the critical mode is again a polariton. 
Instead, when $\omega_c$ equals $\omega_a$, the critical coupling $\eta_c$ vanishes (see supplementary) making the critical mode purely atomic and we refer to this instability as the atomic instability (AI).

Both the OSR and AI critical modes break continuous time-translation invariance and can thus potentially signal a continuous time-crystal phase. However, whether the latter is stable is determined by non-linear effects not included so far. 
In order to capture these in the present interacting many-body system, we perform a systematic perturbative expansion in the relative distance from the critical point $\eta=\left(\tilde{\eta}-\eta_c\right)/\eta_c$.
The resulting effective non-linear equation is of the Stuart-Landau form (see e.g. \cite{Kuramoto1984book}), and is an equation of motion for the collective degrees of freedom which are excited in the SR phases. 
These degrees of freedom constitute the so-called center manifold and are defined by the critical mode, which is composed of both cavity modes as well as of the zero and recoil momentum components of the BEC, $\psi_0$ and $\psi_{\pm Q}$. 
Within the center manifold and to leading order in $\eta$ the recoil momentum component is given by
\begin{equation}\label{eq:OSRrecoil}
    \psi_{\pm Q}(t)=\sqrt{\eta}R\left(c_+ \text{e}^{i\omega_c t}+c_-\text{e}^{-i\omega_c t}\right),
\end{equation}
with $c_\pm$ being the atomic components of the critical-mode eigenvector obtained from the linear analysis \cite{JohansenThesisChp4}.
The cavity fields have the same form with $c_\pm$ replaced by the cavity components of the critical mode. 
Finally, since to leading order the only occupied atom components are $\psi_0$ and $\psi_{\pm Q}$, these are linked by normalization such that
\begin{equation}\label{eq:OSRQ0}
    \psi_0=\sqrt{1-\left|\psi_Q\right|^2-\left|\psi_{-Q}\right|^2}\sim b_0+b_+ \text{e}^{i2 \omega_c t}+\bar{b}_+\text{e}^{-i2\omega_c t},
\end{equation}
with $b_0=1-\eta R^2 \left(\left|c_+\right|^2+\left|c_-\right|^2\right)$ and $b_+=-\eta R^2c_+\bar{c}_-$.
The perturbative approach yields an equation of motion for the SR amplitude, $R$:
\begin{equation}\label{eq:amplitudeEQ}
    \dot{R}=\gamma R- g^r R^3,
\end{equation}
where $\gamma$ is the exponential growth rate of the critical mode obtained from the linear analysis, which in this case can be shown to be positive (see supplementary). 
The non-linearity of the center manifold or in other words, the strength of the self-interaction of the excitations present in the critical mode, is quantified by $g^r$,  (see supplementary for closed expressions for these quantities).
For stable time-crystalline and static solutions, $R$ must be time-independent, real, and positive:
\begin{equation}\label{eq:R}
    R=\sqrt{\frac{\gamma}{g^r}}>0.
\end{equation}
As $\gamma>0$, our analytic solutions can only be stable if $g^r>0$. This is physically clear since otherwise the attractive self-interaction would lead to a first order transition into a phase that requires higher-order non-linearities for stabilization.

The sign of $g^r$ is shown in \cref{fig:lcPD}(b). 
If $\omega_c$ is pushed to $\omega_a$, $g_r=0$ i.e. the self-interaction vanishes as the critical mode is purely atomic, which corresponds to the white region in \cref{fig:lcPD}(b). 
As the fraction $\gamma/g^r$ goes to zero as $\omega_c$ approaches $\omega_a$ (see supplementary), the AI phase has no stable time-crystaline solution.

Short-range interactions between the atoms qualitatively modify $g^r$ and lead to two separatrices in \cref{fig:lcPD}(b). 
The expression for the separatrix $U_{\rm c2}(\Delta)$, drawn with a solid line is given in the supplementary material, while the separatrix $U_{\rm c1}(\Delta)$ between the white and the blue region, is defined by the condition that the energy cost of a Bogoliubov excitation, $\omega_a$, equals $\Delta$. 
When $U>U_{\rm c1}$ the self-interactions of the critical mode become finite and repulsive as $\omega_c<\omega_a$, leading to a finite cavity component of the critical mode.

It is further remarkable that the sign of the self-interactions can be changed via $U$. 
Indeed, within the blue region in \cref{fig:lcPD}(b), that is, for $U_{\rm c1}<U<U_{\rm c2}$, the self-interactions of the critical mode are attractive: $g_r<0$. 
This is due to the fact the short-range repulsion $U$, which penalizes density modulations and in particular excitation of the recoil component $\psi_{\pm Q}$, is not sufficient to counteract the decrease of energy due to coupling to the negatively detuned cavity mode.
The resulting instability of the stationary OSR solution corresponds to a subcritical Hopf bifurcation \cite{KuznetsovBook} of \cref{eq:OSRQ0}. 
On the other hand, when $U>U_{\rm c2}$ (green region in the figure), the short-range repulsion penalizes density-modulations enough to change the sign of the self-interaction of the critical mode and thus stabilize the OSR phase. 
This corresponds to a transition from a subcritical to a supercritical Hopf bifurcation. 

\begin{figure}
    \centering
    \includegraphics[width=0.8\columnwidth]{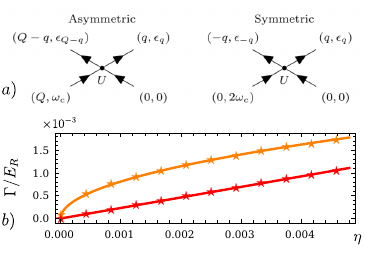}
    \caption{a) The dynamic nature of the OSR phases combined with finite atom-interaction leads to occupation of atom modes out of the center manifold, through the symmetric and asymmetric process illustrated here. b) The scaling of the growth rates, computed from the Floquet quasi-energies of the linearized equations, for the asymmetric channel marked with orange stars, with a square-root fit (orange line) and the scaling of symmetric channel marked with red stars, with a linear fit (red line). The same parameters as in \cref{fig:melting} have been used.}
    \label{fig:scaling}
\end{figure}

\emph{Energy redistribution and melting of the time crystal.---}
The OSR time crystal is thus, up to this point, found to exist in a stable fashion as a supercritical Hopf bifurcation. Still, to fully assess its stability, one must allow for energy redistribution between all degrees of freedom, including those not belonging to the critical polariton mode defining the center manifold of the bifurcation.  We will refer to those as the not-center-manifold (NCM) modes. 
Hence, one needs to treat the many-body problem of scattering between quasi-particles and a time-dependent coherent field.
 
Let us first predict which NCM modes initially participate in the scattering process, assuming we are only slightly into the OSR phase. 
In this regime, we can exploit our analytical knowledge from \cref{eq:OSRrecoil,eq:OSRQ0}.
The fastest-growing NCM mode results from scattering between the atomic components $b_0$ and $c_\pm$ of the center manifold, as illustrated in \cref{fig:scaling}(a). 
For this process, the outgoing NCM modes with energies $\epsilon_q,\epsilon_{q'}$ have to satisfy
$
q+q'=Q,\;\epsilon_q+\epsilon_{q'}=\omega_c.    
$
Since here $q\neq -q'$, we call this the asymmetric channel. Near the critical point, we can approximate $\epsilon_q$ with the Bogoliubov dispersion of the BEC excitations in the absence of the cavity field, which for small $U$ reads $\omega_B(k)\approx E_R k^2+U$. 
This yields $q=Q/2+\sqrt{\omega_c-E_R/2-2U}/\sqrt{2}$ and $q'=Q-q$. 
From the solution of \cref{eq:OSRrecoil,eq:OSRQ0}, we predict an exponential growth of these two Bogoliubov modes with a rate proportional to $U\sqrt{\eta}$.
This asymmetric channel can be closed off if $\omega_c<E_R/2-2U$, which leaves us with a different channel where the component $b_0$ scatters with $b_+$, or $c_+$ with $c_-$. Both these processes produce a symmetric NCM pair with $q=-q'=\sqrt{\omega_c-U}$. One representative process of this symmetric channel is shown in \cref{fig:scaling}(a). In contrast to the asymmetric counterpart, we predict an exponential growth rate proportional to $U \eta$. 

\begin{figure}
    \centering
    \includegraphics[width=\columnwidth]{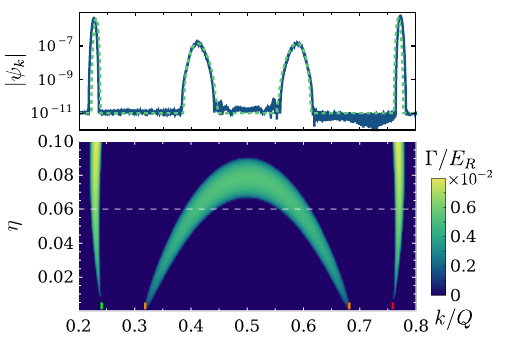}
    \caption{The lower plot shows the exponential growth rates of the atomic modes outside of the center manifold. The parameters are equivalent to those in \cref{fig:lcPD} with $\Delta=0.6E_R$ resulting in $\omega_c=0.586E_R$, and we choose $U=0.01E_R$. The orange ticks indicate the predicted momentum based on the asymmetric channel, while the red tick signifies the symmetric channel momentum. The green tick is the atom mode coupled to the symmetric channel through the cavity. The upper plot shows the resulting atom distribution after 200 periods at the dashed line in the lower plot, both with numerical integration of \cref{eq:atomEOM,eq:cavityEOM} in blue and from the linearized prediction with the dashed green line.}
    \label{fig:melting}
\end{figure}

In order to further verify the above predictions, we have linearized the \cref{eq:atomEOM,eq:cavityEOM} around the OSR phase and extracted the rate by computing the Floquet quasi-energies.
The result of this calculation is shown in the lower panel of \cref{fig:melting}. 
It is seen that the predicted momentum (orange marks for the asymmetric channel and red mark for the symmetric channel) is only reliable close to the phase transition as the dispersion of the NCM mode is quickly modified due to the growing oscillating density modulation.
We also find an additional momentum component that grows (marked in green), which arises from the scattering between a negative momentum NCM mode in the symmetric channel and the recoil component of the center manifold. 
The computed growth rates for the symmetric and asymmetric modes are shown in \cref{fig:scaling}(b), and in both cases, an excellent agreement with our simple predictions based on \cref{fig:scaling}(a) is demonstrated.
Finally, in order to fully confirm our predictions, we performed a full numerical integration using a Runge-Kutta-4 routine, starting from the OSR phase at $\eta=0.06$, corresponding to the white dashed line in the lower panel of \cref{fig:melting}. After evolving the system for 200 periods we compared the momentum distribution with the predictions based on the Floquet quasi-energies and found excellent agreement, as shown in the upper panel of \cref{fig:melting}.

An important outcome of our analysis is that the time crystal is always metastable due to energy redistribution caused by scattering out of the center manifold. Its lifetime, however, increases significantly by considering $\omega_c<E_R/2$ to prohibit the asymmetric scattering processes that lead to much higher growth rates. 

\emph{Conclusions.---}
We have provided a systematic analysis of the role of short-range interactions on the nature and stability of continuous time crystals in dissipative many-body systems of ultracold bosonic atoms in cavities.

First, we have shown that short-range interatomic interactions can alter the nature of the time crystal by transforming the underlying classical bifurcation from sub- to supercritical.

Second, we have studied the effect of short-range interactions on heating and melting of the time crystal.
The heating mechanism we have discussed arises due to the oscillating nature of the atomic fields $\psi_0$ and $\psi_{\left|Q\right|}$. 
As shown in the supplementary material, the amplitude of these fields is not dependent on the details of the underlying critical polaritonic mode, but rather only on the frequency of the oscillations and the proper dimensionless distance from the critical point.

Furthermore, we find that the cavity losses cannot efficiently cool the system \cite{piazza2015self,gambetta2019,lazarides2020time} (NCM modes can be de-excited only at higher order in our expansion, see supplementary material). 
This suggests that the heating mechanism we identified is generic for these cavity systems \cite{Chitra2015}, as long as the cavity line width is comparable to the recoil energy.
We note that time-dependent Hartree-Fock approximations would miss this heating \cite{Zhu2019}, as they lack collisions and thus redistribution \cite{Weidinger2017}. 
As we show it is precisely these effects that eventually lead to the metastable nature of the time-crystalline state, consistent with numerical predictions in related models \cite{Molignini2018,Cosme2022}.

Finally, we point out that the heating mechanism described here is analogous to preheating in the early universe \cite{Kofman1994, Khlebnikov1996}, where the weakly interacting and oscillating, coherent inflation field decays into a cascade of exponentially growing fluctuations, leading to extreme non-equilibrium conditions inaccessible to perturbative methods and finally to prethermalization \cite{Kofman2008}. The analytic discussion presented here corresponds to the linearized classical regime \cite{Boyanovsky1996}, which at later times will be superseded by increasingly non-linear effects leading to a cascade of even more quickly growing fluctuations that eventually thermalize \cite{Micha2003} and thus destroy the time-crystalline phase. 
It will be interesting to pursue this analogy deeper into the highly excited regime using appropriate atom-photon diagrammatic approaches \cite{lang2016critical,lang2020nonequilibrium}.

\begin{acknowledgments}
CHJ would like to thank Johnathan Dubois for many helpful and insightful discussions.
\end{acknowledgments}
\bibliography{LC_biblio}
\newpage
\onecolumngrid
\appendix
\section{Supplementary Material}

\subsection{Center manifold coefficients}
The considered system with two cavity modes is a simplification of the $N$ cavity mode system discussed in \cite{JohansenThesisChp4}. This thesis contains a detailed analysis of the origin of the limit cycle, which we use as foundation for our exploration. 

The starting point is the equations of motion in \cref{eq:cavityEOM,eq:atomEOM}, which we use to define the autonomous system of non-linear first order ODE's
\begin{equation}
\dot{\textbf{v}}=F(\textbf{v}),
\end{equation}
Here $\textbf{v}$ is a vector containing the two complex cavity fields and all the complex atom fields with the different discretized momenta. 
As both cavity modes transfer the same longitudinal momentum ($Q$) and the BEC is initially homogeneous, the emerging critical mode only contains the cavity fields, the homogeneous atom state and the $\pm Q$ atom modes.  These modes constitute the center manifold 
\begin{equation}
\textbf{v}_{cm}=\left(\phi_0,\phi_1,\psi_0,\psi_Q,\psi_{-Q}\right)^T.
\end{equation}
The normal phase $\textbf{X}_{0}=\left(0,0,1,0,0\right)^T$ constitute a fixed point of $F$. 
When $\tilde{\eta}<\eta_c$ this fixed point is stable while it is unstable for $\tilde{\eta}\geq \eta_c$.

As stated in the main text, slightly past the critical point the symmetry-broken state can be approximated as 
\begin{equation}
\textbf{u}=\sqrt{\mu} R \textbf{v}^R \text{e}^{i \omega_c t}+c.c.\,,
\end{equation}
where $\mu=\tilde{\eta}-\eta_c$ is the absolute distance to the critical point and $\omega_c$ is the frequency of the unstable eigenvector $\textbf{v}^R$.  
We will write the equations of motion in terms of the real ($x_\alpha$) and imaginary part ($p_\alpha$) of the complex fields in which the center manifold is spanned by vectors of the form
\begin{equation}
\textbf{v}^R=\left(\textbf{v}^R_{c_1},\textbf{v}^R_{c_2},\textbf{v}^R_{Q},\textbf{v}^R_{-Q}\right)^T,
\end{equation}
with $\textbf{v}^R_\alpha=\left(x_\alpha,p_\alpha\right)^T$. 
This solution is only a good approximation of the new fixed point if the bifurcation is of the supercritical form, which means that the self-interaction of the critical mode is repulsive. 
The linear coefficient in the amplitude equation \cref{eq:amplitudeEQ} is given by the real part of
\begin{equation}\label{eq:lambda}
\lambda=\textbf{v}^L \frac{\partial L}{\partial \mu} \textbf{v}^R,
\end{equation}
where $L=\left.\nabla F\right\vert_{\textbf{X}_0}$ is the Jacobian matrix evaluated at the normal-phase fixed point $\textbf{X}_0$ and $\textbf{v}^L$ ($\textbf{v}^R$) is the left (right) critical eigenvector. We define the linear coefficient as $\gamma=\text{Re}\left(\lambda\right)$. The cubic coefficient of \cref{eq:amplitudeEQ} is given by the real part of
\begin{equation}
g=-\frac{1}{2}\sum_{i,j,k,q}\textbf{v}^L_i \left.\frac{\partial ^3 F_i}{\partial \textbf{X}^j\partial \textbf{X}^k\partial \textbf{X}^q}\right\vert_{\textbf{X}_0,\mu=0} \textbf{v}^R_j\textbf{v}^R_k\bar{\textbf{v}}^R_q=-\textbf{v}^L_i \left(N_0\right)_i^{j,k,q} \textbf{v}^R_j\textbf{v}^R_k\bar{\textbf{v}}^R_q,
\end{equation}
where the same notation as in the main text has been used $g^r=\text{Re}\left(g\right)$. 
Within the center manifold there is no contribution to $g$ from $\partial^2 F$ because the center manifold obeys a reflection symmetry which originates from the fact that the coupled \cref{eq:atomEOM,eq:cavityEOM} posses a $\mathcal{Z}_2$-symmetry as they are invariant under the simultaneous phase shift of the atoms by $\psi_k\rightarrow \text{e}^{i\pi k/Q}\psi_k$ and the cavity fields $\phi_j\rightarrow -\phi_j$. 

\subsubsection{The critical eigenvector}
To compute $\lambda$ and $g$ we use that the right and left eigenvectors are related by 
\begin{equation}\label{eq_LRrelation}
\textbf{v}^L_\alpha=\pm \sigma_x \textbf{v}^R_\alpha,
\end{equation}
where $\sigma_x$ is the first Pauli spin-1/2 matrix. Furthermore, the eigenvectors are normalized such that \cref{eq_LRrelation} is realized with the upper sign. 
The two effective interaction parameters can now be written solely in terms of the right eigenvectors
\begin{equation}
\begin{gathered}
\lambda=\left(\mathds{1}_4\otimes \sigma_x \right)\textbf{v}^R \frac{\partial L}{\partial \mu} \textbf{v}^R,\\
g=-\left(\mathds{1}_4\otimes \sigma_x \right)\textbf{v}^R N_0\textbf{v}^R\textbf{v}^R\bar{\textbf{v}}^R.
\end{gathered}
\end{equation} 
The cavity eigenvector components are connected to the atomic eigenvector components through the definition of the critical eigenvector
\begin{equation}\label{eq_critJac}
\left(L_0-i\omega_c\right)\textbf{v}^R=\textbf{0},
\end{equation}
which leads to the relation
\begin{equation}
\textbf{v}^R_{c_j}=-\sqrt{2}\eta_c \beta_j\mqty(1&0&1&0\\\frac{\kappa+i \omega_c}{\Delta_j}&0&\frac{\kappa+i \omega_c}{\Delta_j}&0)\mqty(x_Q\\p_Q\\x_{-Q}\\p_{-Q}),
\end{equation}
where 
\begin{equation}
\beta_j= \frac{\Delta_j}{2} \frac{\Delta_j^2+\kappa^2-\omega_c^2-2i\omega_c\kappa}{\Delta_j^4+2\Delta_j^2\left(\kappa^2-\omega_c^2\right)+\left(\kappa^2+\omega_c^2\right)^2}
\end{equation}
and the critical coupling, derived in \cite{JohansenThesisChp4}, is given by
\begin{equation}\label{eq:etac}
    \eta_c=\sqrt{\frac{\omega_a^2-\omega_c^2}{E_R\sum_{j=1,2}\frac{\Delta_j \left(\Delta_j^2+\kappa^2-\omega_c^2\right)}{\omega_c^4+2\omega_c^2\left(\kappa^2-\Delta_j^2\right)+\left(\Delta_j^2+\kappa^2\right)^2}}}.
\end{equation}
From the critical eigenvalue condition $\det\left(L_0-I\omega_c\right)=0$ one finds
\begin{equation}\label{eq_betaRelations}
\begin{aligned}
\eta_c^2\beta&=\eta_c^2\sum_{j=1,2} \text{Re}\beta_j=\frac{\omega_a^2-\omega_c^2}{2E_R},\\
\text{Im}\beta_j&=0.
\end{aligned}
\end{equation} 
Linearizing around the normal phase means that the short-range interaction only couples the modes $Q$ and $-Q$ in a symmetric manner. 
As the cavity also couples identically to these two modes, the components of the critical eigenvector obeys $x_Q=x_{-Q}=x_a$ and $p_Q=p_{-Q}=p_a$. 
Using this symmetry $x_a$ and $p_a$ can be connected through \cref{eq_critJac} and one finds
\begin{equation}\label{eq_paRelation}
p_a=\frac{i\omega_c}{E_R}x_a=i\tilde{\omega} x_a,
\end{equation}
where the dimensionless frequency $\tilde{\omega}=\omega_c/E_R$ has been introduced for later convenience. 
Now $\textbf{v}^R$ can be fully expressed through the parameters of our theory and $x_a$. 
A closed-form expression for $x_a$ can be found through the normalization condition
\begin{equation}\label{eq_xa}
\textbf{v}^L\textbf{v}^R=1\rightarrow x_a=\frac{1}{2}\left(4\eta_c^2\sum_j\left[\beta_j^2\frac{\kappa+i\omega_c}{\Delta_j}\right]+\frac{i\omega_c}{E_R}\right)^{-1/2}.
\end{equation}
This form of $x_a$ guarantees the upper sign in \cref{eq_LRrelation}. 

\subsubsection{Computing $g^r$ and $\gamma$}
By substituting the  critical eigenvector into \cref{eq:lambda} one finds the expression
\begin{equation}\label{eq_lambda}
\lambda=\frac{8}{\eta_c}x_a^2\eta_c^2 \beta=\frac{2\eta_c^2\beta}{\eta_c\left(4\eta_c^2\sum_j\left[\beta_j^2\frac{\kappa+i\omega_c}{\Delta_j}\right]+\frac{i\omega_c}{E_R}\right)}=\frac{\omega_a^2-\omega_c^2}{E_R}\frac{1}{\eta_c\left(4\eta_c^2\sum_j\left[\beta_j^2\frac{\kappa+i\omega_c}{\Delta_j}\right]+\frac{i\omega_c}{E_R}\right)}.
\end{equation}
The expression for $g$ is 
\begin{equation}\label{eq_g}
\begin{aligned}
g&=x_a^2\abs{x_a}^2E^R\Bigg(\tilde{U}\left(3+2\tilde{\omega}^2+3\tilde{\omega}^4\right)+4\left(1-\tilde{\omega}^2\right)\left(3+\tilde{\omega}^2\right)\Bigg)\\
&=x_a^2\abs{x_a}^2E_R W_a\left(\tilde{U},\tilde{\omega}\right)
\end{aligned}
\end{equation}
where the dimensionless interaction is defined as $\tilde{U}=U/E_R$.

It is clear that the only part that makes both $\lambda$ and $g$ complex is in $x_a^2$.  As the coefficients for our theory are related to the real part of $\lambda$ and $g$, it is relevant to extract the real part of $x_a^2$
\begin{equation}\label{eq_xa2}
\begin{aligned}
\text{Re}\left(x_a^2\right)&=\eta_c^2\kappa\frac{\sum_j\frac{\beta_j^2}{ \Delta_j}}{\abs{\left(4\eta_c^2\sum_j\left[\beta_j^2\frac{\kappa+i\omega_c}{\Delta_j}\right]+\frac{i\omega_c}{E_R}\right)}^2}\\&=\frac{\eta_c^2 \kappa}{2\abs{\left(4\eta_c^2\sum_j\left[\beta_j^2\frac{\kappa+i\omega_c}{\Delta_j}\right]+\frac{i\omega_c}{E_R}\right)}^2}\sum_j\Delta_j \frac{\left(\Delta_j^2+\kappa^2-\omega_c^2\right)^2+4\omega_c^2\left(\Delta_j^2-\omega_c^2\right)}{\left(\Delta_j^4+2\Delta_j^2\left(\kappa^2-\omega_c^2\right)+\left(\kappa^2+\omega_c^2\right)^2\right)^2}.
\end{aligned}
\end{equation}
This directly shows that the only dependence on $U$ in $x_a$ is through $\eta_c$ in \cref{eq:etac}.
Due to the complexity of the full closed form expression of $g$ it is insightful to consider the behavior of $\text{Re}\left(x_a^2\right)$ and $W_a$ separately. 

First considering $W_a$
\begin{equation}
W_a\left(\tilde{U},\tilde{\omega}\right)=\tilde{U}\left(3+2\tilde{\omega}^2+3\tilde{\omega}^4\right)+4\left(1-\tilde{\omega}^2\right)\left(3+\tilde{\omega}^2\right).
\end{equation}
The interesting feature of $W_a$ is the fact that it has a sign change through a zero-crossing at a critical frequency $\tilde{\omega}_0$ such that
$W_a\left(\tilde{U},\tilde{\omega}_0\right)=0$.
The closed form expression for $\tilde{\omega}_0$ is
\begin{equation}\label{eq_omega0}
\tilde{\omega}_0=\sqrt{\frac{\tilde{U}-4+2\sqrt{2}\sqrt{8-4\tilde{U}-\tilde{U}^2}}{4-3\tilde{U}}}=\sqrt{1+\frac{\tilde{U}}{2}+\mathcal{O}(\tilde{U}^2)}.
\end{equation}

This exactly defines the separatrix $U_{c2}$ shown as a black line in \cref{fig:lcPD}(b).
\cref{fig:lcPD}(b) is plotted as a function of $\Delta$ and not $\omega_c$ because of the atom instability. 
In the regime where $\kappa<E_R$, the equations simplify because near $\Delta\sim E_R$ one has that $\omega_c\approx\Delta$. 
This means that one can replace $\tilde{\omega}_0$ with $\Delta/E_R$ instead of substituting in the full expression in \cref{eq:frequency}. 
The relevant quantity that one should compare $\tilde{\omega}_0$ to is the dimensionless frequency of the bare atomic instability, which happens at
\begin{equation}
\tilde{\omega}_a=\sqrt{1+2\tilde{U}},
\end{equation}
and which sets the dashed separatrix $U_{c1}$ in \cref{fig:lcPD}(b).
For $\tilde{U}=0$ the frequencies $\tilde{\omega}_0$ and $\tilde{\omega}_a$ coincide, which means that there will be no cubic interactions for the atomic instability without short-range interactions.
As $\tilde{U}$ is made finite we see from the expansion in \cref{eq_omega0} that $\tilde{\omega}_a>\tilde{\omega}_0$ for small $\tilde{U}<1$.  
By keeping the full expression for $\tilde{\omega}_0$, one finds that the critical $\tilde{U}_c$ where $\tilde{\omega}_a=\tilde{\omega}_0$ is 
\begin{equation}\label{eq_Uc}
\tilde{U}_c=\sqrt{2},
\end{equation}
which is the intersection point of the separatrices at finite $U$ with $\Delta=\sqrt{1+2\sqrt{2}}E_R$.
Below this interaction strength, $\tilde{\omega}_0$ is smaller than $\tilde{\omega}_a$. 
The effect is that $W_a(\tilde{U},\tilde{\omega}_a)<0$ for all $\tilde{U}<\tilde{U}_c$.

To determine the nature of the interactions one has to determine the sign of $\text{Re}\left(x_a^2\right)$. 
This sign is fixed by the numerator of \cref{eq_xa2} and using the parametrization discussed in the main text one finds
\begin{equation}\label{eq_signxa2}
\sum_{j=1,2}\Delta_j \left(\left(\Delta_j^2+\kappa^2-\omega_c^2\right)^2+4\omega_c^2\left(\Delta_j^2-\omega_c^2\right)\right)=\frac{ \delta}{2}\left(\frac{\left(\delta^2+4\kappa^2\right)^2}{16}+\Delta^2\left(3\delta^2+4\kappa^2\right)+8\Delta^4\right).
\end{equation}
So for any values of $\kappa$, $\Delta$ and $\abs{\delta}$, the sign of $\text{Re}\left(x_a^2\right)$ is set by the sign of $\delta$.
This means that for a chosen sign of $\delta$ the sign of $W_a$ determines whether the non-linear self-interactions are repulsive or attractive.   
Additionally this also means that $\gamma>0$. If $\delta>0$ then $\omega_c<\omega_a$ and both $\eta_c^2\beta$ in \cref{eq_betaRelations} and $\text{Re}\left(x_a^2\right)$ are greater than zero. If $\delta<0$ then $\omega_c>\omega_a$ and both $\eta_c^2\beta$ and $\text{Re}\left(x_a^2\right)$ are negative such that $\gamma$ is again positive. 

Next consider the fraction $\gamma/g^r$ which determines the magnitude of the stable time-crystalline phase. By using the above derived relations one can show that it scales with $\sqrt{\epsilon}$ in the limit where $\omega_c^2\rightarrow \omega_a^2-\epsilon$ with $\epsilon\ll \{E_R,\Delta_{1/2},\kappa\}$
\begin{equation}
\begin{aligned}
   \lim_{\omega_c^2\rightarrow \omega_a^2-\epsilon} \frac{\gamma}{g^r}&=\lim_{\omega_c^2\rightarrow \omega_a^2-\epsilon}\frac{8}{\eta_c}\frac{\omega_a^2-\omega_c^2}{2E_R}\frac{\text{Re}\left(x_a^2\right)}{\text{Re}\left(x_a^2\right)\abs{x_a}^2E_R W_a\left(U/E_R,\omega/E_R\right)}\\&=\lim_{\omega_c\rightarrow \omega_a}\frac{32}{\eta_c}\frac{\omega_a^2-\omega_c^2}{2E_R^2 W_a\left(U/E_R,\omega/E_R\right)}\left\vert 4\eta_c^2\sum_j\left[\beta_j^2\frac{\kappa+i\omega_c}{\Delta_j}\right]+\frac{i\omega_c}{E_R}\right\vert\\
    &=\lim_{\omega_c^2\rightarrow \omega_a^2-\epsilon}\frac{32}{\sqrt{\frac{\omega_a^2-\omega_c^2}{E_R\sum_{j=1,2}\frac{\Delta_j \left(\Delta_j^2+\kappa^2-\omega_c^2\right)}{\omega_c^4+2\omega_c^2\left(\kappa^2-\Delta_j^2\right)+\left(\Delta_j^2+\kappa^2\right)^2}}}}\frac{\omega_a^2-\omega_c^2}{2E_R^2 W_a\left(U/E_R,\omega/E_R\right)}\left\vert 4\eta_c^2\sum_j\left[\beta_j^2\frac{\kappa+i\omega_c}{\Delta_j}\right]+\frac{i\omega_c}{E_R}\right\vert\\
    &=32 \sqrt{\epsilon}\frac{\sqrt{E_R\sum_{j=1,2}\frac{\Delta_j \left(\Delta_j^2+\kappa^2-\omega_a^2\right)}{\omega_a^4+2\omega_a^2\left(\kappa^2-\Delta_j^2\right)+\left(\Delta_j^2+\kappa^2\right)^2}}}{2E_R^2 W_a\left(U/E_R,\omega/E_R\right)}\frac{\omega_a}{E_R}+\mathcal{O}\left(\epsilon^{3/2}\right).
    \end{aligned}
\end{equation}
This is important as it proves that the AI region does not possess a stable time-crystalline solution, to leading order in $\mu$, as $\epsilon\rightarrow 0$ for the AI. 

The fact that we have analytical expressions for all the important quantities also allows us to show some intriguing features of the time-crystalline phase within the center manifold. The first important feature was discussed in the main text, namely that the frequency of OSR phase is independent of the atom parameters. The second important feature we will show now is that the time-averaged occupation in the recoil field is only indirectly depending on the cavity parameters. As stated in the conclusions, this is means that our heating discussion is is more generic, as it does not depend on the specific cavity configuration.
If we write \cref{eq:OSRrecoil} using $x_a$ and $p_a$ the occupation in the recoil mode is given by
\begin{equation}
\begin{aligned}
     \abs{\psi_Q(t)}^2&=\frac{1}{2}\left(\abs{x_a}^2+\abs{p_a}^2\right)\\&=\frac{\mu}{4}R^2\left(2\abs{x_a}^2+x_a^2\exp(i 2 \omega_c t)+\bar{x}_a^2\exp(-i 2\omega_c t)+2\abs{p_a}^2+p_a^2\exp(i 2 \omega_c t)+\bar{p}_a^2\exp(-i 2\omega_c t)\right).
\end{aligned}
\end{equation}
Due to the periodicity of the system the time average is given by
\begin{equation}\label{eq_tavgPsi}
    \left<\abs{\psi_Q}^2\right>_T=\int_0^{2\pi/\omega_c} \abs{\psi_Q(t)}^2\dd t=\frac{\mu}{2}R^2\left(\abs{x_a}^2+\abs{p_a}^2\right)=\frac{\mu}{2}R^2\abs{x_a}^2\left(1+\tilde{\omega}^2\right),
\end{equation}
Where $p_a$ have been eliminated through \cref{eq_paRelation}. 
Using the results from \cref{eq_lambda,eq_g} we find 
\begin{equation}
    R^2=\frac{\text{Re}\left(\lambda\right)}{\text{Re}\left(g\right)}=\frac{4\left(\tilde{\omega}_a^2-\tilde{\omega}^2\right)}{\eta_c \abs{x_a}^2 W_a\left(\tilde{U},\tilde{\omega}\right)}.
\end{equation}
Inserting this into \cref{eq_tavgPsi} we find 
\begin{equation}\label{eq_atomInvariance}
    \left<\abs{\psi_Q}^2\right>_T=\frac{2 \eta \left(\tilde{\omega}_a^2-\tilde{\omega}^2\right) \left(1+\tilde{\omega}^2\right)}{W_a\left(\tilde{U},\tilde{\omega}\right)},
\end{equation}
which only depends on the atom parameters, the OSR frequency, and the relative depth into the OSR phase, $\eta$. 
The same $\tilde{\omega}$ can be generated with many different cavity configurations, for example by changing $\delta$ and having a small $\kappa$ or even more generally by departing from the fully symmetric case presented here. 

\subsection{Including fluctuations outside the center manifold}
While our theory within the center manifold predicts that the time crystal is stable also with finite interactions $U$, it does not capture atom modes outside of the center manifold.  
The contact interaction allows occupation in the center manifold to scatter to the other atom modes with momenta different from $\pm Q$ and 0. 
Inside the OSR phase the NCM modes can be occupied due to the presence of the OSR. 
This leads to heating and potentially also the destruction of the time crystal in the long time limit. 
One way to describe this is to linearize around the OSR solution $\textbf{v}_\text{osr}(t)$
\begin{equation}
\textbf{v}(t)=\textbf{v}_\text{osr}(t)+\delta\textbf{v}(t).
\end{equation}
This leads to an equation for the fluctuations
\begin{equation}
\begin{gathered}
\dot{\textbf{v}}=\dot{\textbf{v}}_\text{osr}+\dot{\delta\textbf{v}}=F(\textbf{v}_\text{osr}+\delta\textbf{v})=F(\textbf{v}_\text{osr})+\left.\nabla F\right|_{\textbf{v}=\textbf{v}_\text{osr}}\delta\textbf{v}+\mathcal{O}\left(\delta\textbf{v}^2\right)\\
\rightarrow \dot{\delta\textbf{v}}=\left.\nabla F\right|_{\textbf{v}=\textbf{v}_\text{osr}}\delta\textbf{v}+\mathcal{O}\left(\delta\textbf{v}^2\right)\approx J_\text{osr}(t)\delta\textbf{v},
\end{gathered}
\end{equation}
where $J_\text{osr}(t)$ is a time-dependent matrix-valued function. 
Using the approximate fixed point from the analytical OSR solution we can derive an approximate form of $J_\text{osr}(t)$. 
Due to the periodicity of the OSR solution $J_{\text{osr}}(t)$ can be expanded in a discrete Fourier series of the form
\begin{equation}
J_\text{osr}(t)=\sum_{n=-4}^4 M_n \text{e}^{i n \omega_{c} t}. 
\end{equation}
The coupling to the cavity in \cref{eq:atomEOM} is proportional to a product of a NCM mode and a cavity field. 
To first order in fluctuations there is therefore no coupling between cavity fluctuations and the NCM modes.
For the leading-order heating mechanism $\delta \textbf{v}$ only includes the atom modes with momentum $k\notin\{0,Q,-Q\}$ and is therefore solely described by \cref{eq:atomEOM} with the cavity fields replaced by the OSR solution. 
It is for this reason that we are able to use the simple scattering description, discussed in the main text, to predict the momentum of the growing modes. 

Because the cavity loss has already been used to stabilize the OSR phase within the center manifold this means that the cavity is not able to cool down the NCM modes at the linear level.
As one includes higher orders in fluctuations the cavity fluctuations can potentially start cooling down the NCM but as this is a higher-order effect, fine tuning would be needed to make it overcome the first-order heating before the system has thermalized and the OSR phase is destroyed.

To verify our simple scattering predictions we derive $J_\text{osr}(t)$ from \cref{eq:atomEOM}. The linearized equation for the NCM mode with momentum $k$ is
\begin{equation}\label{eq_linearized_NCM}
\begin{aligned}
    i\partial_t\psi_k=&\left(-\dot{\Omega}+k^2-U\abs{\hat{\psi}_0}^2+2U\left(\abs{\hat{\psi}_0}^2+\abs{\hat{\psi}_Q}^2+\abs{\hat{\psi}_{-Q}}^2\right)\right)\psi_k+U\left(\hat{\psi}_0^2+2\hat{\psi}_{-Q}\psi_{Q}\right)\bar{\hat{\psi}}_{-k}\\&+\frac{\tilde{\eta}}{\sqrt{2}}\sum_{j}\Re\left(\hat{\psi}_j\right)\left(\psi_{k+Q}+\psi_{k-Q}\right),
\end{aligned}
\end{equation}
where the hat has been used to identify the OSR components that are approximated as unchanged within the linearization. The overall phase of the atoms is set by $\dot{\Omega}$ and chosen such that $\Im{\psi_0}=0$ within the center manifold \cite{JohansenThesisChp4}. Within the linearization the value is 
\begin{equation}
\begin{aligned}
    \dot{\Omega}=&\frac{\tilde{\eta}}{\sqrt{2}}\sum_{j}\Re(\phi_j)\frac{\psi_Q+\bar{\psi}_Q+\psi_{-Q}+\bar{\psi}_{-Q}}{2\psi_0}\\&+U\left(2-\psi_0^2+\frac{1}{2}\left[\left(\psi_Q+\bar{\psi}_{Q}\right)\left(\psi_{-Q}+\bar{\psi}_{-Q}\right)+\left(\psi_Q-\bar{\psi}_{Q}\right)\left(\psi_{-Q}-\bar{\psi}_{-Q}\right)\right]\right).
    \end{aligned}
\end{equation}
From \cref{eq_linearized_NCM} we see that the finite occupation of the cavity field leads to coupling of the $k$ NCM mode with the NCM mode at $k\pm Q$. As the occupation of the NCM fields are small and we consider $\omega_c<E_R$, one can truncate after only one recoil kick such that $\abs{k}<Q$. This is confirmed by the full numerical solution of \cref{eq:atomEOM,eq:cavityEOM} shown in the main text. With this truncation each NCM, $\psi_k$, only couples to the seven other fields $\{ \bar{\psi}_k,\,\psi_{-k},\,\bar{\psi}_{-k},\,\psi_{k-Q},\,\bar{\psi}_{k-Q},\,\psi_{-k+Q},\,\bar{\psi}_{-k+Q}\}$.
For each value of $k$ we therefore find a $J_\text{osr}(t)$ given by
\begin{equation}
    J_\text{osr}(t)=i\mqty(-m_k&0&0&-g_k&-m_Q&-g_Q&0&0\\0&\bar{m}_k&\bar{g}_k&0&\bar{g}_Q&\bar{m}_Q&0&0\\0&-g_k&-m_k&0&0&0&-m_Q&-g_Q\\\bar{g}_k&0&0&\bar{m}_k&0&0&\bar{g}_Q&\bar{m}_Q\\-m_Q&-g_Q&0&0&-m_{k-Q}&0&0&-g_k\\\bar{g}_Q&\bar{m}_Q&0&0&0&\bar{m}_{k-Q}&\bar{g}_k&0\\0&0&-m_Q&-g_Q&0&-g_k&-m_{k-Q}&0\\0&0&\bar{g}_Q&\bar{m}_Q&\bar{g}_k&0&0&\bar{m}_{k-Q}),
\end{equation}
with the vector $\delta\textbf{v}^T=(\psi_k,\, \bar{\psi}_k,\,\psi_{-k},\,\bar{\psi}_{-k},\,\psi_{k-Q},\,\bar{\psi}_{k-Q},\,\psi_{-k+Q},\,\bar{\psi}_{-k+Q})^T$ and the five different entries being
\begin{equation}\label{eq_Jelements}
    \begin{aligned}
    m_k&=k^2+U-2U\left(\hat{\psi}_Q^2+\bar{\hat{\psi}}_Q^2\right)-\frac{\tilde{\eta}}{\sqrt{2}}\sum_{j}\Re(\hat{\phi}_j)\frac{\hat{\psi}_Q+\bar{\hat{\psi}}_Q}{\hat{psi}_0},\\
    m_{k-Q}&= m_{k\rightarrow k-Q},\\
    m_{Q}&=\frac{\tilde{\eta}}{\sqrt{2}}\sum_{j}\Re(\hat{\phi}_j)+2U\hat{\psi}_0\left(\bar{\hat{\psi}}_Q+\hat{\psi}_Q\right),\\
    g_k&=U\left(2\hat{\psi}_Q^2+\hat{\psi}_0^2\right),\\
    g_Q&=2U\hat{\psi}_Q\hat{\psi}_0.
    \end{aligned}
\end{equation}
Inserting the OSR solutions into \cref{eq_Jelements} one finds an analytical expression for $J_\text{osr}(t)$ which is periodic such that $J_\text{osr}(t)=J_\text{osr}(t+T)$ with $T=2\pi/\omega_c$. We then employ standard Floquet theory by numerically time-evolving the eight equations over one period $T$. This allows us to find the fundamental matrix $\Phi(t)$ which is defined as the solution to
\begin{equation}
    \partial_t \Phi(t)=J_\text{osr}(t) \Phi(t),
\end{equation}
with the initial condition $\Phi(0)=\mathds{1}_8$. The eigenvalues $\lambda_i$ of the monomdromy matrix $M=\Phi(T)$ determines the growth rates of the NCM modes $\Gamma_i=\Re(\log(\lambda_i)/T)$. To understand the initial heating effects we only need to investigate the eigenmode with the largest growth rate $\Gamma=\max(\Gamma_i)$.
By Computing $\Gamma$ as a function of $k$ we are able to compute the growth rates of the different channels as plotted in \cref{fig:melting}.

\end{document}